\documentclass[reprint,aps,ams,amsmath,prx,twocolumn,superscriptaddress,longbibliography]{revtex4-1}
\usepackage{graphics}
\usepackage{graphicx}
\usepackage{epsfig}
\usepackage{amsmath}
\usepackage{amssymb}
\usepackage{amsfonts}
\usepackage{bm}
\usepackage{mathrsfs}
\usepackage{color}
\usepackage{textcomp}
\usepackage[T1,T2A]{fontenc}
\usepackage[utf8]{inputenc}
\usepackage{textcomp}
\usepackage{xcolor}
\usepackage{bbm}

\usepackage[normalem]{ulem}
\usepackage{comment}
\usepackage{soul}

\newcommand{\be}{\begin{equation}}
\newcommand{\ee}{\end{equation}}
\def\rr#1{(\ref{#1})}
\newcommand{\lambdaD}{\lambda_\mathrm{D}}
\newcommand{\mrm}[1]{\mathrm{#1}}

\newcommand{\taud}{\tau_\mathrm{D}}
\newcommand{\taurck}{\tau_{\mathrm{RC},k}}

\renewcommand{\Im}{\operatorname{Im}}

\newcommand{\hrho}{\widehat{\rho}}

\usepackage{hyperref}
\begin{document}
\title{Universal behavior in diffuse charge dynamics with patterned electrodes}

\author{E. Krucker-Velasquez}
\affiliation{Center for Computation and Theory of Soft Materials, Robert R. McCormick School of Engineering and Applied Science, Northwestern University, Evanston IL 60208 USA
}
\author{E. Kirkinis}
\affiliation{Center for Computation and Theory of Soft Materials, Robert R. McCormick School of Engineering and Applied Science,  Northwestern University, Evanston IL 60208 USA
}
\affiliation{Department of Materials Science \& Engineering, Northwestern University, Evanston IL 60208 USA}

\author{M. Olvera de la Cruz}
\affiliation{Center for Computation and Theory of Soft Materials, Robert R. McCormick School of Engineering and Applied Science,  Northwestern University, Evanston IL 60208 USA
}
\affiliation{Department of Physics and Astronomy, Northwestern University, Evanston, IL 60208 USA}
\affiliation{Department of Materials Science \& Engineering, Northwestern University, Evanston IL 60208 USA}

\date{\today}

\begin{abstract}
Iontronic devices that replicate physiological processes will improve energy efficiency to support artificial-intelligence architectures. Energy storage and conversion devices, however, display disordered interfacial and bulk configurations that conventional impedance spectroscopy cannot resolve. Here, we show that the onset of frequency dispersion in the impedance of an electrolytic cell with \emph{non-uniformly charged electrodes} takes place at approximately twice the resonator natural frequency, independently of whether the cell is driven by oscillating charges or voltage differences, and remains valid for both overlapping and non-overlapping double layers. In addition to characterizing energy storage devices, these \emph{universal} findings apply to devices interfaced with biomaterials. 
\end{abstract}

\maketitle
The rapid growth of investment in artificial intelligence architectures is accompanied by an implicit expectation for either greater energy production or more efficient use of existing resources. The latter has renewed interest in aqueous ionic systems, where low-voltage ion transport can be combined with physiologically compatible biodegradable substrates, and chemically reconfigurable architectures \cite{Chun2015}. Alongside vertical and lateral mermsistors, where information is carried by electrons and holes in a crystal lattice \cite{Sangwan2020}, aqueous architectures offer a complementary route in which ions respond to both electrical and chemical signals.

Realizing this potential requires understanding how the structure and heterogeneity of the interface shape ionic transport and the resulting electrical response. Porosity, roughness, variations in surface chemistry, and nonuniform current distributions introduce a spectrum of local resistive and capacitive responses in the electrodes of energy-storage and energy-conversion devices. Standard descriptions such as the constant-phase element \cite{deLevie1963}, commonly used to represent the resulting frequency dispersion of nonideal interfaces \cite{Jorcin2006,*CordobaTorres2015,*Gateman2022}, compress the spatial complexity of the interface into a distribution of relaxation times and, in consequence, do not uniquely identify the physical origin of the dispersion nor determine how a prescribed surface-charge pattern reorganizes the neighboring electrolyte. 
The need for a spatially resolved description extends beyond energy-storage and energy-conversion devices. Impedance measurements are also used to characterize biological materials and can, for example, distinguish benign from malignant tissue through differences in their measured electrical properties \cite{Halter2008}. Yet lateral charge heterogeneity, which is also an intrinsic and sometimes functional feature of biological interfaces, is rarely accounted for. \\
\begin{figure*}
\includegraphics[width=0.98\linewidth]{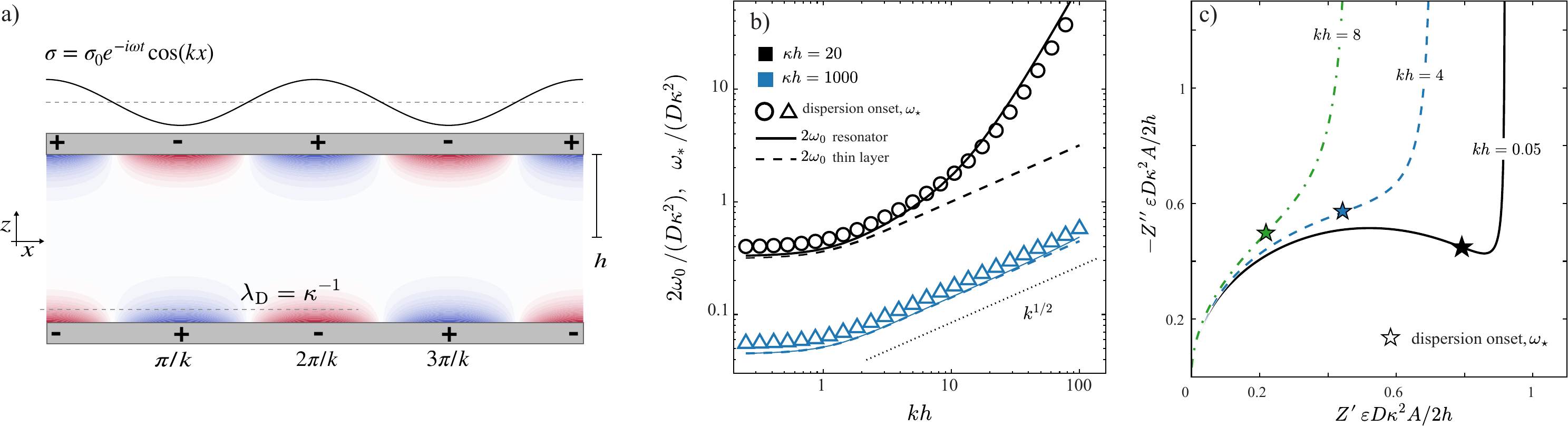}
\caption{\label{fig:Fig1}(a) Electrolytic cell with inhomogeneous wall charge distribution of wavenumber $k$; the equilibrium double layer screens the pattern over a width $\lambda_D=\kappa^{-1}$. (b) The dispersion-onset frequency $\omega_\star$ (markers) is approximated well by twice the natural frequency $\omega_0$ Eq. \rr{omega0}  of the cell in the resonator interpretation (solid lines) - see also Eq. \rr{omega0app}. $\omega_0\sim k^{1/2}$ is the theoretical prediction \rr{omega0kh}. (c) Nyquist representation of the impedance \rr{Z} of the cell displayed in panel (a). The starred markers denote the onset of frequency dispersion $\omega_\star$, which occurs at an inflection point of the Nyquist plot. }
\end{figure*}
In this Letter, we introduce a \emph{universal} frequency governing the impedance dispersion of an electrolytic cell with \emph{nonuniformly charged} walls. Its universality is reflected in the onset of dispersion, which, across pattern wavelength and confinement, occurs at twice the natural frequency $\omega_0$:
\be \label{omega0}
\omega_0^2 =\frac{1}{2} (D \kappa^2) (Dk\kappa) \coth kh,
\ee
where $k$ is the wall charge modulation wavenumber, $D$ the species diffusion coefficient, $\kappa$ the inverse Debye length and $2h$ the cell width, cf. Fig. \ref{fig:Fig1}(a). The variation of (approximately twice) the natural frequency \rr{omega0} with respect to the electrode charge modulation wavenumber $k$ is displayed in Fig. \ref{fig:Fig1}(b). In the same plot we superpose the threshold $\omega_\star$ at which frequency dispersion sets in, defined to occur at a frequency giving rise to \emph{an inflection point} in the Nyquist plot of the impedance 
\be \label{Z}
Z= -\frac{2h}{A} \frac{
1 - \frac{D\kappa^2}{i\omega } \frac{\tan Ph}{Ph} \frac{kh}{\tanh kh}}{ \epsilon D (P^2 + k^2)  },
\ee
cf. Fig. \ref{fig:Fig1}(c).
Eq. \rr{Z} is derived in the Supplementary Material addendum from the Poisson-Nernst-Planck equations in the Debye-Falkenhagen approximation \cite{Bazant2004,Shrestha2025},
where $\epsilon$ is the permitivity of the electrolyte, $\omega$ the driving frequency, $P = (i\omega /D -\kappa^2 - k^2)^{1/2}$ a complex wavenumber determining both the penetration of the charge into the electrolyte and its phase delay relative to the wall charge, and $A$ is the cell's cross-sectional area.  Frequency dispersion sets-in when the impedance departs from its ideal capacitor scaling $Z\sim (i\omega)^{-1}$, represented by a vertical straight line in the Nyquist plot. 

 The three markers in Fig. \ref{fig:Fig1}(c) mark the Nyquist inflection points. The frequency-dispersion threshold, $\omega_\star$, decreases monotonically with wall-charge modulation wavelength, consistent with the trend reported in the numerical simulations of \citet{Alexander2016}. 

\begin{figure}[b!]
\vspace{-5pt}
\begin{center}
\includegraphics[width=0.98\linewidth]{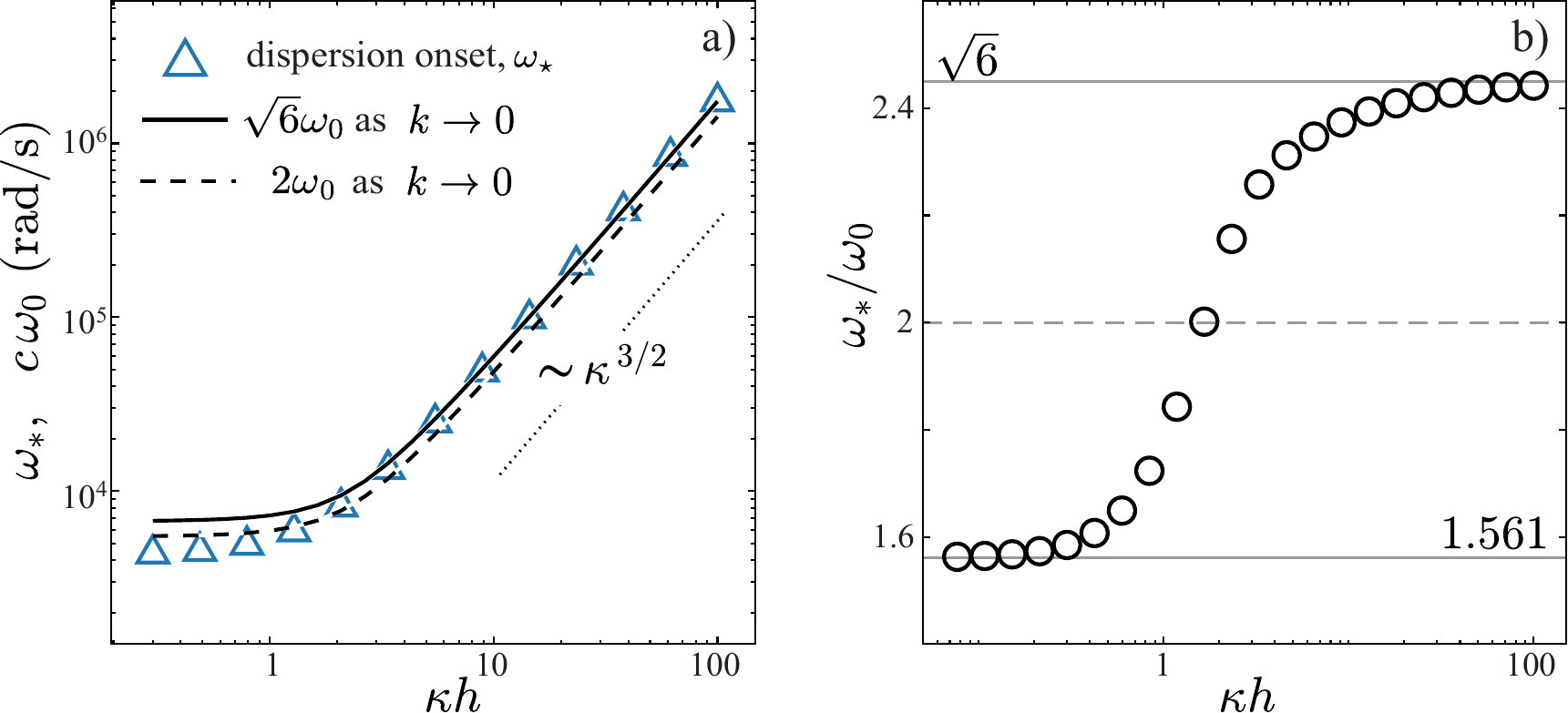}
\vspace{-20pt}
\end{center}
\caption{Uniformly-charged electrodes ($k\rightarrow 0$): 
(a) Twice the natural frequency, $2\omega_0$, from \rr{omega0q0} (solid line), closely approximates the threshold for the onset of frequency dispersion, $\omega_\star$, (triangles), defined by the Nyquist inflection point of the impedance \rr{Z} in the limit $k\rightarrow 0$. The agreement holds for both overlapping and non-overlapping Debye layers. The asymptotics \rr{omega0kh} are clearly visible. Here, $h$ is 1 micron wide.
(b) Frequency ratio, $\omega_\star/\omega_0$,
bounded as described in Eq. \rr{ratio}.
\label{Zomega_kappah}  }
\vspace{-10pt}
\end{figure}

The natural frequency $\omega_0$ has two distinguished limits, 
\be \label{omega0kh}
\omega_0 \sim \frac{1}{\sqrt{2}}
\left\{
\begin{array}{cc}
\displaystyle
 D\kappa^{3/2} k^{1/2}, & kh\gg1\\
\displaystyle
D\kappa^{3/2} h^{-1/2}, & kh\ll 1\\
\end{array}
\right. \; .
\ee
The $\omega_0 \sim k^{1/2}$ scaling for $k h\gg1$ is easily appreciable in Fig. \ref{fig:Fig1}(b). In contrast, the $kh\ll1$ limit recovers the uniform-cell result, with its corresponding $\sim\kappa^{3/2}$ scaling shown in Fig. \ref{Zomega_kappah}(a). 

\begin{figure*}[t!]
\includegraphics[width=0.75\linewidth]{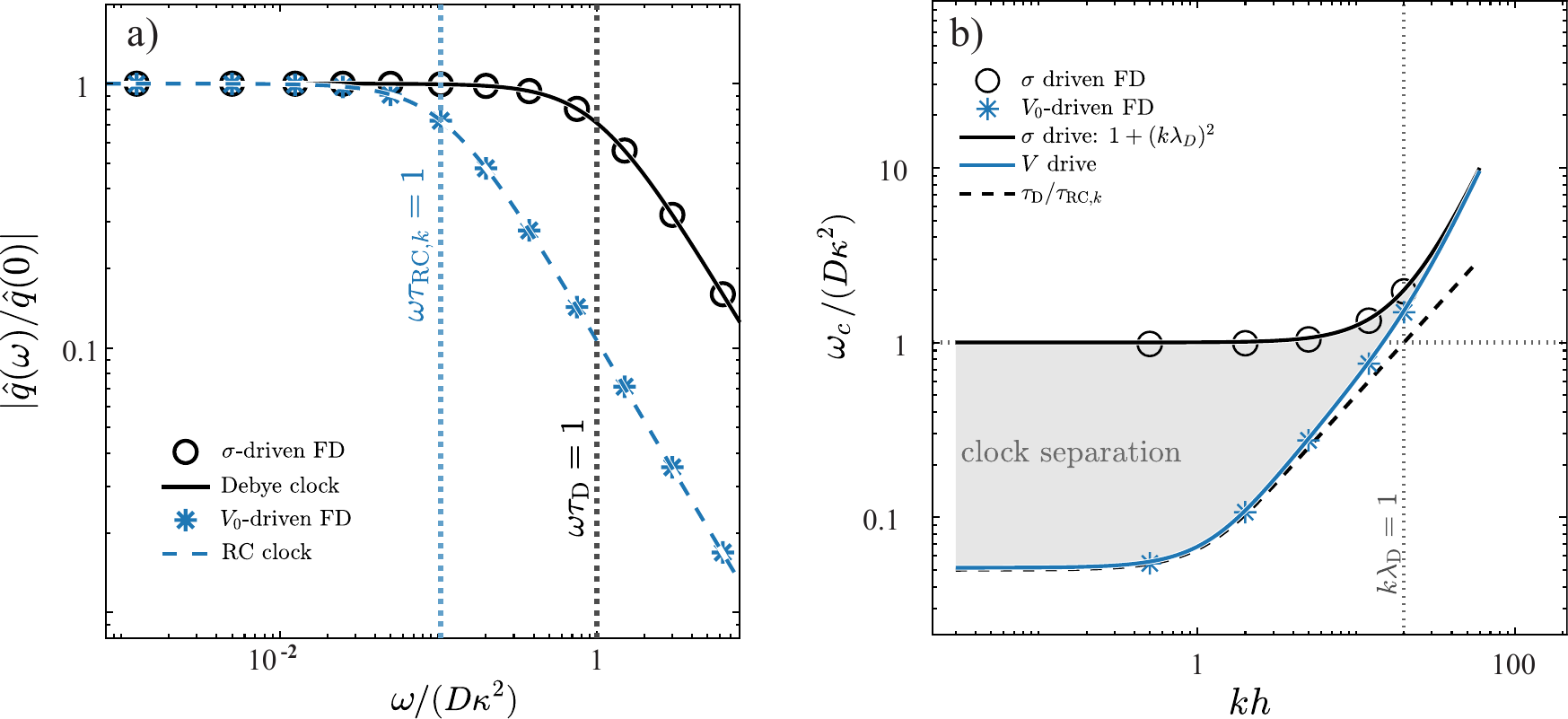}
\caption{\label{fig:Fig2} Pattern-dependent diffuse-layer charge and relaxation: a) diffuse-layer charge averaged over the half width $h$. The solid and dashed lines show the exact inear response under charge and voltage regulation, respectively; markers show results from the same time-domain simulations that produce the indistinguishable impedances in Fig.~\ref{fig:Fig1}(b). Blue and black vertical dotted lines mark the RC and Debye relaxation scales, respectively. (b) Characteristic relaxation frequency $\omega_c\taud$ of the diffuse-layer charge, defined by $|\hat q(\omega_c)|=|\hat q(0)|/\sqrt{2}$ under charge (black) and voltage (blue) regulation. Solid lines correspond to the crossovers as derived in the SM.  The charge-regulated frequency obeys Eq.~\eqref{eq:omegac} exactly, whereas the voltage-regulated frequency follows the RC prediction $\omega_c\taud=\taud/\taurck$ but departs from it as $k\lambdaD\to1$.}
\end{figure*}

The relation between the dispersion threshold, $\omega_\star$, and the natural frequency, $\omega_0$, can also be established analytically. In the thin-layer limit, $\kappa\to \infty$ (but keeping the ratio $\omega/(D\kappa^2)$ constant), the signed curvature $|\partial_\omega Z\times \partial^2_\omega Z|$ (treating the impedance as a three-dimensional vector) changes signs at $\omega_\star^2 = 3(D\kappa^2)(Dk\kappa) \coth kh $, which, compared to Eq. \rr{omega0} gives 
\be \label{freqratioinf}
\omega_ \star / \omega_0 = \sqrt{6}\sim 2.45,
\ee
cf. the Supplementary Material (SM) addendum for the derivation. The thin Debye layers limit in Eq. \rr{freqratioinf} sets the upper bound of the ratio $\omega_\star/\omega_0$. In panel (b) of Fig. \ref{Zomega_kappah} we show the numerically-determined ratio for uniformly-charged walls, which varies with Debye-layer thickness within
\be 
\label{ratio}
1.561<\omega_\star/\omega_0\leq\sqrt{6}\; .
\ee
The natural frequency thus sets the dispersion scale from overlapping to thin Debye layers. The middle horizontal line, $\omega_\star/\omega_0=2$, shows that $2\omega_0$ provides a useful approximation to the threshold across this range.

The natural frequency $\omega_0$ can be derived (cf. End Matter \ref{sec: QF}) by considering the 
quality factor $Q$ \citep{Sader2010,*Burg2007,*Burg2009} 
\be \label{Q}
{Q} =\omega \frac{ \bar{\mathscr{E}}}{ \bar{\dot{\mathscr{E}}}}
\ee
for the oscillatory aqueous ionic system of Fig. \ref{fig:Fig1}(a),
where $\omega$ is the driving frequency, $ \bar{\mathscr{E}}$ is the energy stored by the fields and $\bar{\dot{\mathscr{E}}}$ the power, both observables averaged over the period of oscillation (denoted by a bar).
The frequency asymptotics of $Q$ (cf. \rr{Qlow}) lead to
\be \label{Qas}
Q \sim 
\left\{
\begin{array}{cl} \displaystyle
\frac{Dk\kappa \coth kh}{4\omega} & \textrm{as } \omega \rightarrow 0 \; (\textrm{and } \kappa \rightarrow \infty), 
\vspace{8pt}\\
\displaystyle
\frac{\omega}{2D\kappa^2} & \textrm{as } \omega \rightarrow\infty.
\end{array}
\right.
\ee
The asymptotics \rr{Qas} are those of a driven oscillator \cite{Puri1987} of damping coefficient $2r = D\kappa^2$ and 
natural frequency \rr{omega0} (cf. End Matter \ref{sec: osc}). 

The two branches of $Q(\omega)$ in \rr{Qas} characterize distinct charging regimes controlled by the time scales 
\be
\tau_D = (D\kappa^2)^{-1} \; \textrm{and} \; 
\taurck=(D k \kappa \coth(kh))^{-1}.
    \label{timescales}
\ee
At low frequency, the applied field varies slowly enough for the ions to form nearly quasistatic diffuse layers adjacent to the charged walls. Electric energy is then stored with little dissipation per cycle relative to the stored energy. As the frequency increases, the diffuse charge develops a small phase delay relative to the wall charge, even while local Debye relaxation remains fast, $\omega\taud\ll1$. The resulting incomplete screening permits the penetrating field, although weak, to extend beyond the diffuse layer. In the thin-layer regime, $\lambdaD\ll\ell_k=\tanh(kh)/k$, this field acts over a much larger region than the near-wall field and can therefore store comparable energy despite its smaller amplitude. To leading order, the ratio of penetrating- to screened-field energies is $2\omega^2\taud\taurck$. The natural-frequency scale \rr{omega0} $\omega_0\approx(2\taud\taurck)^{-1/2}$ therefore corresponds to equal energies in these two field components to leading order and lies near the minimum of the quality factor. At high frequency, $\omega\taud\gg1$, diffuse-layer charging is suppressed, and the penetrating field dominates electrostatic storage. Displacement current dominates the response, while ionic conduction remains dissipative. The shorter forcing period reduces the loss per cycle relative to stored energy, giving $Q\sim\omega\taud/2$.\\
In contrast, the Nyquist inflection reflects the competition between the bulk and diffuse layer contributions to the impedance.  In the thin layer intermediate regime $\Im \tilde{Z}\approx \omega\taud + 1/(\omega\taurck)$. The bulk contribution grows with frequency, while the diffuse layer decreases with frequency.  Its reversal occurs at a higher frquency $\omega_\star =\sqrt{6}\omega_0$ where the bulk contribution is approximately three times that of the diffuse layer. 

In the distinguished limit $kh\ll 1$, Eq.~\rr{omega0kh} identifies an intermediate relaxation time, $\tau_0\sim h^{1/2}\lambda_D^{3/2}/D$ which is the geometric mean of the Debye relaxation time, $\taud=\lambda_D^2/D$, and the blocking-electrode charging time, $\tau_{\mathrm{RC}}=h\lambda_D/D$. In the usual thin-double-layer limit, $\kappa h\gg1$, these times are ordered as  $\tau_{\mathrm D}<\tau_0<\tau_{\mathrm{RC}}$. The same intermediate scale was identified by \citet[Eq.~(14)]{Rubinstein2009} through the reactance
extremum of a blocking electrolytic cell. Our uniform-cell limit recovers this scale: for $\kappa h\gg1$, the imaginary part of Eq.~\eqref{Z} is extremal, to leading order, at $\omega\approx(\taud\tau_{\mathrm{RC}})^{-1/2}
=\sqrt{2}\omega_0$. We note that $\omega_\star$ is larger than the frequency found by the reactance extremum. The reactance extremum is a first-derivative condition where the gap-averaged ionic and displacement currents contribute equally to the out-of-phase current. This equality does not yet mark the loss of diffuse-layer control: although its reactance is already smaller than the bulk contribution, the ionic response continues to determine inflection point in the Nyquist plot. The curvature reverses only at $\omega_\star\simeq\sqrt{6}\omega_0$, a factor $\sqrt{3}$ above the reactance minimum, when the displacement current is three times the ionic contribution. In the overlapping-double-layer regime, $\kappa h<1$, the ordering is reversed, $ \tau_{\mathrm{RC}}<\tau_0<\tau_{\mathrm D}$, although $\tau_0$ remains intermediate between the two characteristic times. This ordering also determines whether the corresponding mode is strongly damped. Since $\omega_0\sim\tau_0^{-1}$ and the damping coefficient is $2r=D\kappa^2$. For $\kappa h\gg1$, the damping rate exceeds the natural frequency, and the response is overdamped. When $\kappa h\lesssim1$, the Debye layers associated with the opposing walls overlap and $\omega_0$ becomes comparable to, or larger than, the damping rate, allowing the response to become oscillatory. 

The discussion above identifies the Debye and RC times as the two limiting clocks of diffuse-layer charging. We now show that this clock dichotomy is a coarse-pattern effect: it persists when $k^{-1}\gg\lambda_D$, but disappears as the lateral pattern scale approaches the Debye length. Although voltage- and charge-driven configurations share the same impedance, Eq.~\eqref{Z}, their diffuse-layer relaxation dynamics are distinct. For homogeneous electrodes, the diffuse double layer is already known to relax on protocol-dependent timescales. The relaxation spectra of the diffuse-charge, $\hat{q}(\omega)=\int_0^h\hrho(\omega)\mrm{d}z$, shown in Figure \ref{fig:Fig2} a) depict that this protocol-dependence is also present in the patterned cell. The difference in crossover frequencies is a result of the electric fields available to assemble the screening charge under the different drives. Under charge regulation (open black markers in Fig.~\ref{fig:Fig2}), the prescribed wall charge is present from the outset, and ions within a Debye length experience the full unscreened field. The diffuse layer therefore relaxes on the Maxwell timescale, $\tau_{\mathrm{D}}\sim\lambda_D^2/D$. Under voltage regulation, by
contrast, the patterned-cell response (blue star symbols in Fig.~\ref{fig:Fig2}) is governed by the geometric field $V_0/\ell$. For homogeneous electrodes, $\ell\approx h$. Here, however, surface patterning introduces an additional dependence on the lateral wavenumber $k$, such that $\ell\approx\min(h,k^{-1})$. Because the field lines close through the bulk between neighboring regions of the surface pattern, storing $\varepsilon V_0/\lambda_D$ against that weaker field takes $\ell\lambda_D/D$, leads to
\begin{equation}
  \frac{\taurck}{\taud}\approx \kappa \min\left(h,k^{-1}\right),
  \label{eq:pathratio}
\end{equation}
which interpolates between the uniform-cell time $\lambdaD h/D$ \cite{Bazant2004} as $kh\to0$ and the electrode-array time $\lambdaD/(Dk)$ \cite{Ramos1999,Ajdari2000} as $kh\to\infty$.  The diffuse-charge spectra in Fig.~\ref{fig:Fig2}(a) yield the characteristic relaxation frequencies shown in Fig.~\ref{fig:Fig2}(b).  \\
\indent Figure~\ref{fig:Fig2}(b) follows the diffuse-charge relaxation frequencies, where we define the characteristic frequency $\omega_c$ to satisfy $|\hat q(\omega_c)|=|\hat q(0)|/\sqrt{2}$, under charge-, $\omega_c^\sigma$, and voltage-, $\omega_c^V$, regulation across $kh$. For the single-pole-approximation of the $\hat{q}(\omega)$ the relaxation frequencies are
\begin{equation}
    \omega_c^\sigma = D(\kappa^2 + k^2)\quad \textrm{and} \quad\omega_c^V\approx \taurck^{-1}.
    \label{eq:omegac}
\end{equation}
As made explicit by Figure~\ref{fig:Fig2}(b),  patterning compresses the ratio of distances, and at the screening scale, $k\lambdaD\to1$, erases the distinction between a fixed-charge surface and a wired electrode. We note that neither clock is an equivalent-circuit time constant: $\taud$ and $\taurck$ govern the assembly of the diffuse charge cloud itself, which a lumped $R$-$C$ representation of the interface does not resolve.

\begin{figure}
\vspace{-5pt}
\begin{center}

\includegraphics[width=0.9\linewidth]{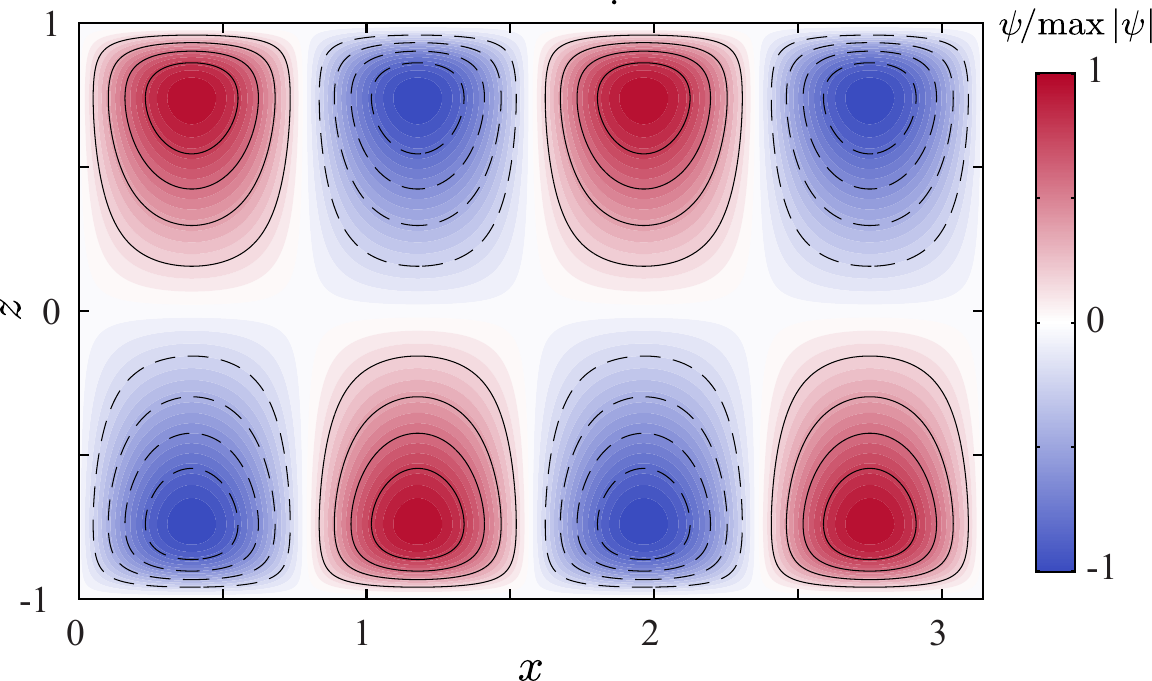}
\vspace{-10pt}
\end{center}
\caption{\label{fig:psi_total}
Hydrodynamic flow induced in the charging cell of Fig. \ref{fig:Fig1}(a): Numerically-determined flow streamlines with superposed streamfunction normalized amplitude. Adjacent vortices 
display an opposite sense of circulation and 
oscillate with a period of $2\pi/\omega$ seconds. }
\vspace{0pt}
\end{figure}

Whereas the incompressibility condition precludes the establishment of flow in a charging cell with uniformly-charged electrodes, this is not the case when the electrodes are patterned. In this case, the system looses its translational symmetry and forms a periodic group of counter-rotating oscillatory vortices. The
streamfunction $\psi = \psi(x,z, t)$ for a two-dimensional incompressible liquid satisfies 
\be \label{psinl}
\rho_l\partial_t \nabla^2 \psi = \eta \nabla^4 \psi  + \partial_x \rho \partial_z \phi - \partial_z\rho \partial_x \phi,
\ee
where $\rho_l$ and $\eta$ are the liquid mass density and viscosity and
$\psi(x, \pm h, t) = 0 = \partial_z \psi(x,\pm h, t)$ (the no-slip boundary condition) and 
$ 
\mathbf{v} = (u, 0, w) = \left(\partial_z \psi, 0, -\partial_x \psi \right)$ is the liquid velocity. 
In Fig. \ref{fig:psi_total} we display the flow pattern induced in the charging cell of Fig. \ref{fig:Fig1}(a), determined numerically. It is composed of rows of counter-rotating vortices (whose sense of circulation is denoted by the solid/dashed lines). This flow, generated at zero Pecl\'et number, does not affect the predictions of the theory developed in this Letter. When the Pecl\'et number becomes significant, the bulk charge will evolve relative to an advective current. This can somewhat affect, for instance, the value of the resonator natural frequency, but the character of the results described in this Letter remain unchanged. 

The values of the electrolytic cell damping coefficient and natural frequency \rr{omega0q0} discussed above, fall within the range of published parameters fitting experimental measurements of complex conductivity for saline solutions of KCl, $\textrm{MgCl}_2$,
$\textrm{CaCl}_2$ and NaCl to a harmonic restoring potential model \cite{Sanabria2006,KruckerVelasquez2025}. For instance, when $2h=10^{-3}$ m and $\kappa = 10^7 \textrm{m}^{-1}$, the damping coefficient is $D\kappa^2\sim 10^5 \;\textrm{sec}^{-1}$ and the natural frequency $\omega_0 \sim  10^3$ Hz, both values agreeing well with the fitting parameters to the experiment displayed in the right-most two columns of Table II in the above reference. 
\\
\noindent
\textbf{Acknowledgments}\\ 
This work was supported by the Department of Energy (DOE), Office of Basic Energy Sciences under Contract No. DE-FG02-08ER46539.
The authors are grateful to Prof. Mark Orazem for helpful suggestions and to Dr. Ahis Shrestha for providing Ref. \cite{Rubinstein2009}. 
\\\\
\textbf{Declaration of Interests}\\ The authors report no conflict of interest.
\begin{center}
\textbf{End Matter}
\end{center}
\vspace{-0.4in}
\section{\label{sec: QF}Quality factor for the charging cell with inhomogeneously-charged electrodes}
The quality factor $Q$ of Eq. \rr{Q} for the electrolytic cell of Fig. \ref{fig:Fig1}(a) is
\be \label{Qlow}
Q =\omega \frac{ \epsilon \langle \overline{\nabla \phi \cdot \nabla \phi  }\rangle}{ -2\langle \overline{ \mathbf{i} \cdot \nabla \phi } \rangle },
\ee
where $
\mathbf{i} = - D\left[\nabla \rho + \epsilon \kappa^2 \nabla\phi \right] 
$ is the ionic current in the Debye-Falkenhagen approximation  \citep{Bazant2004}, 

$\rho$ is the bulk charge of the electrolyte $\phi$ the electric potential induced in the charging cell of Fig. \ref{fig:Fig1}(a) (cf. SM),
$
\kappa = \left[ \frac{2e^2 c_\infty}{\epsilon k_BT} \right]^{\frac{1}{2}}
$ the inverse Debye length, $e$ the proton charge, $c_\infty$ the uniform undisturbed solute concentration, $\epsilon$ is the permitivity of the liquid and $k_BT$ the thermal energy. 
Angle brackets denote averaging over the cross-section of the cylinder and an overbar denotes time-averaging over the period of oscillation of the field. 
The energy and energy dissipated per unit time are
\be
\frac{1}{2}\nabla \phi \cdot \nabla \phi , \quad
-\mathbf{i} \cdot \nabla \phi = D\left[ \nabla \rho + \epsilon \kappa^2 \nabla \phi  \right]\cdot \nabla \phi 
\ee
(where we remind the reader that $\epsilon \kappa^2 D$ is the conductivity of the medium). 
Employing the identity $ \rho  = \epsilon (k^2 \phi - \phi_{zz})$ they can directly be expressed as 
\begin{widetext}
\be \label{J0a}
\frac{1}{2}\nabla \phi \cdot \nabla \phi = \frac{1}{4}\left[ \phi_z \phi^*_z + k^2 \phi \phi^*\right], \quad 
\mathbf{i} \cdot \nabla \phi =  \frac{D}{4} \left[ (\rho_z \phi^*_z + \rho^*_z \phi_z) + k^2  (\rho \phi^* + \rho^* \phi) 
+ 2\epsilon \kappa^2 \left( \phi_z \phi^*_z + k^2 \phi \phi^* \right)
\right].
\ee
\end{widetext}
Expressions \rr{J0a} are already averaged over the period of oscillation of the external field and only depend on the transverse channel coordinate (the coordinate $z$, cf. Fig. \ref{fig:Fig1}(a)). From now on we discuss only the 
averaged over the channel width energy and energy dissipation, e.g.
\be \label{disav}
-\langle\mathbf{i} \cdot \nabla \phi  \rangle = -\frac{1}{2h} \int_{-h}^h dz \mathbf{i} \cdot \nabla \phi .
\ee
We formulate the quality factor \rr{Qlow} and calculate its asymptotics in the limits $\omega\rightarrow 0$ and $\omega \rightarrow \infty$ which resemble the asymptotics \rr{Qasosc} of a driven damped oscillator. Thus, it is easy to determine 
the natural frequency of oscillation which takes the form
\begin{widetext}
\be\label{omega0app}
\omega_0 = -\frac{2 \kappa^{4} k D^{2} \left(\left(k^{2}+\frac{\kappa^{2}}{2}\right) \sinh \! \left(2 h K\right)+h K\, \kappa^{2}\right) \cosh \! \left(k h \right)}{-2K^3\left( \cosh \! \left(2 h K\right)+1\right) \sinh \! \left(k h \right)+k \left(2 k^{2}+3 \kappa^{2}\right) \cosh \! \left(k h \right) \sinh \! \left(2 h K\right)-2 K \cosh \! \left(k h \right) h \,\kappa^{2} k}
\ee
\end{widetext}
where $K = \sqrt{k^{2}+\kappa^{2}}$. In the limit $\kappa \rightarrow \infty$ the natural frequency \rr{omega0app} acquires its simpler form \rr{omega0}. 

In Fig. \ref{fig:Fig1}(b) we display the the two times the natural frequency $\omega_0$ in Eq. \rr{omega0app} (continuous line) for relatively thick Debye layers. 
It is seen that $2\omega_0$ approximates the threshold of frequency dispersion (triangular markers) quite well. 
The simplified natural frequency \rr{omega0} obtained from \rr{omega0app} by taking the limit $\kappa \rightarrow \infty$, 
scales as $ k^{1/2}$ and is shown with the dashed line of slope $1/2$ in Fig. \ref{fig:Fig1}(b). 

\begin{figure}[t]
\vspace{-10pt}
\centering
\includegraphics[width=0.8\linewidth]{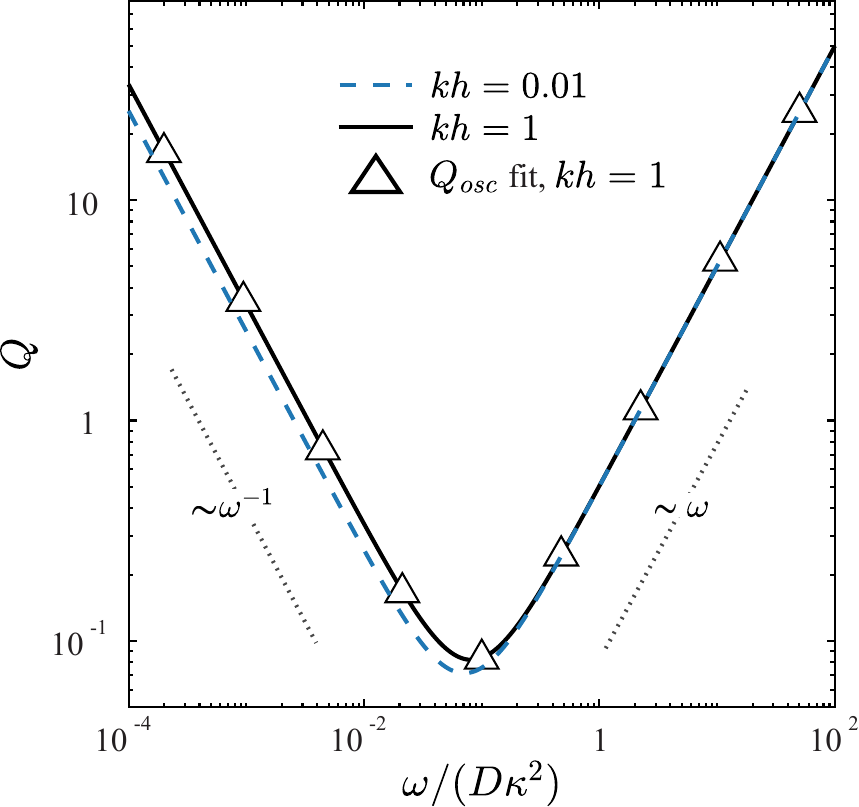}
\vspace{-10pt}
\caption{Quality factor $Q$ Eqs. \rr{Q} \& \rr{Qlow} for the charging cell with inhomogeneously charged electrodes displayed in Fig. \ref{fig:Fig1}(a) at two values of the wall-charge wavenumber $k$. The two dotted lines of slopes $-1$ and $1$ show that the asymptotics of the charging cell are those of a damped driven oscillator, cf. Eq. \rr{Qas}. 
The triangular markers denote the fit
\rr{Qfit} and provide an excellent approximation to the exact quality factor.}\label{QomegaND}
\vspace{-10pt}
\end{figure}

The above discussion leads us to employ the oscillator quality factor (cf. Eq. \rr{Qosc}) here expressed in the form
\be \label{Qfit}
Q_{\textrm{fit}} = \frac{\omega}{2D\kappa^2} \left[ 1 + \left( \frac{\omega_0}{\omega}\right)^2 \right]
\ee
where $\omega_0$ is given by \rr{omega0},
as a faithful fit of its
electrokinetic counterpart. This is displayed in figure \ref{QomegaND} with the triangular markers (for $kh=1$ only). The fit 
curve \rr{Qosc} thus provides an excellent approximation to the true quality factor curves of the same figure. 

The aforementioned main results of this Letter are also applicable in charging cells with uniformly charged electrodes ($k\rightarrow 0$). In this limit, the impedance \rr{Z} recovers a form already known in the literature \cite{Freire2006,*Stout2015}
\be \label{ZB}
Z= -\frac{2h}{A} 
\frac{1 - \frac{D\kappa^2}{i\omega } \frac{\tan Ph}{Ph} }{ \epsilon D P^2 },
\ee
where now $P = (i\omega /D -\kappa^2 )^{1/2}$. 
In Fig. \ref{Zomega_kappah} we display the threshold for frequency dispersion $\omega_\star$ (triangular markers) for various values of the  Debye inverse wavelength $\kappa$. We again define the threshold for the onset of frequency dispersion to occur at a frequency giving rise to an inflection point in the Nyquist plot of impedance \rr{ZB}, which resembles panel (c) of Fig. \ref{fig:Fig1}.

The quality factor \rr{Q} in the uniform wall charge distribution case ($k\rightarrow 0$) gives rise to the natural frequency
\be \label{omega0q0}
\omega_0 \sim \frac{D \kappa^{2} \left[2 h \kappa +\sinh \! \left(2 h \kappa \right)\right]^{1/2}}{\left[2 h \kappa  \cosh \! \left(2 h \kappa \right)+4 h \kappa -3 \sinh \! \left(2 h \kappa \right)\right]^{1/2}}. 
\ee
The continuous curve on the right panel of Fig. \ref{Zomega_kappah} displays the behavior of (twice) $\omega_0$ in \rr{omega0q0}. 
The natural frequency \rr{omega0q0} gives an excellent description of the threshold frequency leading to frequency dispersion (triangular markers) for both overlapping and non-overlapping Debye layers lying on opposite cell walls. 

\section{\label{sec: osc}Quality factor for a forced damped oscillator}
Consider the oscillator
\be \label{osc}
m\ddot{x} + 2rm \dot{x} +m \omega_0^2 x = F \sin \omega t
\ee
where $m$, $r$, $\omega_0$, and $F$ have their usual dimensional interpretations. 
The time-averaged energy $E_{av}$ and power supply $P_{av}$ are \citep{Puri1987}
\be
E_{av} = \frac{1}{4} m \left(\frac{F}{Z}\right)^2  \left[ 1 + \left( \frac{\omega_0}{\omega}\right)^2 \right], \quad P_{av} = \frac{F^2 R}{2 Z^2}, 
\ee
where the mechanical resistance $R$, reactance $X$ and impedance $Z$ are defined by 
\be
R = 2rm, \quad X = m\left(\omega - \frac{\omega_0^2}{\omega} \right), \quad Z^2 = X^2 + R^2. 
\ee
Thus, the quality factor of the oscillator is 
\be \label{Qosc}
Q_{osc} =\omega \frac{E_{av}}{P_{av}} = \frac{\omega}{4r} \left[ 1 + \left( \frac{\omega_0}{\omega}\right)^2 \right].
\ee
The frequency asymptotics of \rr{Qosc} are 
\be \label{Qasosc}
Q_{osc} \sim 
\left\{
\begin{array}{cc}
\frac{1}{4r} \frac{\omega_0^2}{\omega} & \textrm{as } \omega \rightarrow 0, \\
\frac{\omega}{4r} & \textrm{as } \omega \rightarrow\infty.
\end{array}
\right.
\ee

\bibliography{biblio}

@book{Melcher1981,
  author    = {Melcher, James R.},
  title     = {Continuum Electromechanics},
  publisher = {MIT Press},
  address   = {Cambridge, MA},
  year      = {1981},
}

@article{Henrique2025,
  author  = {Henrique, Filipe and Gupta, Ankur},
  title   = {Parallel {W}arburg elements describe ionic transport in nanopores},
  journal = {PRX Energy},
  volume  = {4},
  number  = {2},
  pages   = {023009},
  year    = {2025},
  doi     = {10.1103/PRXEnergy.4.023009},
}

@article{Shrestha2025b,
  author  = {Shrestha, Ahis and Kirkinis, Eleftherios and Olvera de la Cruz, Monica},
  title   = {Self-generated electrokinetic flows from active-charged boundary patterns},
  journal = {Physical Review Research},
  volume  = {7},
  number  = {2},
  pages   = {023223},
  year    = {2025},
  doi     = {10.1103/PhysRevResearch.7.023223},
}

@article{KruckerVelasquez2025,
  author  = {Krucker-Velasquez, Emily and Bazant, Martin Z. and Alexander-Katz, Alfredo and Swan, James W.},
  title   = {Dynamic response of concentrated electrolytes to chirp signals},
  journal = {ACS Nano},
  volume  = {19},
  pages   = {13673--13684},
  year    = {2025},
  doi     = {10.1021/acsnano.4c14099},
}

@article{Stout2015,
  author  = {Stout, Robert F. and Khair, Aditya S.},
  title   = {Moderately nonlinear diffuse-charge dynamics under an ac voltage},
  journal = {Physical Review E},
  volume  = {92},
  number  = {3},
  pages   = {032305},
  year    = {2015},
  doi     = {10.1103/PhysRevE.92.032305},
}

@article{Freire2006,
  author  = {Freire, Fernando C. M. and Barbero, Giovanni and Scalerandi, Marco},
  title   = {Electrical impedance for an electrolytic cell},
  journal = {Physical Review E},
  volume  = {73},
  number  = {5},
  pages   = {051202},
  year    = {2006},
  doi     = {10.1103/PhysRevE.73.051202},
}

@book{Panton1996,
  author    = {Panton, Ronald L.},
  title     = {Incompressible Flow},
  edition   = {2},
  publisher = {John Wiley \& Sons},
  address   = {New York},
  series    = {A Wiley-Interscience Publication},
  pages     = {xv+780},
  year      = {1996},
}

@book{Bluman1989,
  author    = {Bluman, George W. and Kumei, Sukeyuki},
  title     = {Symmetries and Differential Equations},
  publisher = {Springer},
  address   = {New York},
  series    = {Applied Mathematical Sciences},
  volume    = {81},
  year      = {1989},
  doi       = {10.1007/978-1-4757-4307-4},
}

@article{Alexander2016,
  author  = {Alexander, Christopher L. and Tribollet, Bernard and Orazem, Mark E.},
  title   = {Contribution of surface distributions to constant-phase-element ({CPE}) behavior: 2. {C}apacitance},
  journal = {Electrochimica Acta},
  volume  = {188},
  pages   = {566--573},
  year    = {2016},
  doi     = {10.1016/j.electacta.2015.11.135},
}

@article{Halter2008,
  author  = {Halter, Ryan J. and Schned, Alan and Heaney, John and Hartov, Alex and Schutz, Shannon and Paulsen, Keith D.},
  title   = {Electrical Impedance Spectroscopy of Benign and Malignant Prostatic Tissues},
  journal = {The Journal of Urology},
  year    = {2008},
  volume  = {179},
  number  = {4},
  pages   = {1580--1586},
  doi     = {10.1016/j.juro.2007.11.043}
}

@article{Jorcin2006,
  author  = {Jorcin, Jean-Baptiste and Orazem, Mark E. and P{\'e}b{\`e}re, Nadine and Tribollet, Bernard},
  title   = {{CPE} Analysis by Local Electrochemical Impedance Spectroscopy},
  journal = {Electrochimica Acta},
  year    = {2006},
  volume  = {51},
  number  = {8--9},
  pages   = {1473--1479},
  doi     = {10.1016/j.electacta.2005.02.128}
}

@article{CordobaTorres2015,
  author  = {C{\'o}rdoba-Torres, Pedro and Mesquita, Thiago J. and Nogueira, Ricardo P.},
  title   = {Relationship between the Origin of Constant-Phase Element Behavior in Electrochemical Impedance Spectroscopy and Electrode Surface Structure},
  journal = {The Journal of Physical Chemistry C},
  year    = {2015},
  volume  = {119},
  number  = {8},
  pages   = {4136--4147},
  doi     = {10.1021/jp512063f}
}

@article{Gateman2022,
  author  = {Gateman, Samantha Michelle and Gharbi, Ouma{\"i}ma and de Melo, Herc{\'i}lio Gomes and Ngo, Kieu and Turmine, Mirelle and Vivier, Vincent},
  title   = {On the Use of a Constant Phase Element ({CPE}) in Electrochemistry},
  journal = {Current Opinion in Electrochemistry},
  year    = {2022},
  volume  = {36},
  pages   = {101133},
  doi     = {10.1016/j.coelec.2022.101133}
}

@article{deLevie1963,
  author  = {de Levie, R.},
  title   = {On porous electrodes in electrolyte solutions: {I}. {C}apacitance effects},
  journal = {Electrochimica Acta},
  volume  = {8},
  number  = {10},
  pages   = {751--780},
  year    = {1963},
  doi     = {10.1016/0013-4686(63)80042-0},
}

@book{puri1987,
  title={Fundamentals of Vibrations and Waves},
  author={Puri, S.P.},
  year={1987},
  publisher={Tata Mcgraw-Hill Book Comp., New Delhi}
}

@article{Burg2009,
  title={Nonmonotonic energy dissipation in microfluidic resonators},
  author={Burg, T.P. and Sader, J.E. and Manalis, S.R.},
  journal={{Physical Review Letters}},
  volume={102},
  number={22},
  pages={228103},
  year={2009},
  publisher={APS}
}

@article{Burg2007,
  title={Weighing of biomolecules, single cells and single nanoparticles in fluid},
  author={Burg, T.P. and Godin, M. and Knudsen, S.M. and Shen, W. and Carlson, G. and Foster, J.S. and Babcock, K. and Manalis, S.R.},
  journal={Nature},
  volume={446},
  number={7139},
  pages={1066--1069},
  year={2007},
  publisher={Nature Publishing Group UK London}
}

@article{Sader2010,
  title={Energy dissipation in microfluidic beam resonators},
  author={Sader, J.E. and Burg, T.P. and Manalis, S.R.},
  journal={{Journal of Fluid Mechanics}},
  volume={650},
  pages={215--250},
  year={2010},
  publisher={Cambridge University Press}
}

@article{Chun2015,
  title={Iontronics},
  author={Chun, H. and Chung, T. D.},
  journal={{Annual Review of Analytical Chemistry}},
  volume={8},
  number={1},
  pages={441--462},
  year={2015},
  publisher={Annual Reviews}
}

@article{Sangwan2020,
  title={Neuromorphic nanoelectronic materials},
  author={Sangwan, V.K. and Hersam, M.C.},
  journal={Nature Nanotechnology},
  volume={15},
  number={7},
  pages={517--528},
  year={2020},
  publisher={Nature Publishing Group UK London}
}

@article{Shrestha2025,
  title={Universal behaviour in boundary-driven electrokinetic flows},
  author={Shrestha, A. and Kirkinis, E. and Olvera de la Cruz, M.},
  journal={{Journal of Fluid Mechanics}},
  volume={1010},
  pages={A50},
  year={2025},
  publisher={Cambridge University Press}
}

@article{Ehrlich1982,
  title={Bipolar model for traveling-wave induced nonequilibrium double-layer streaming in insulating liquids},
  author={Ehrlich, R.M. and Melcher, J.R.},
  journal={{The Physics of Fluids}},
  volume={25},
  number={10},
  pages={1785--1793},
  year={1982},
  publisher={American Institute of Physics}
}

@article{Ajdari2000,
  author  = {Ajdari, Armand},
  title   = {Pumping liquids using asymmetric electrode arrays},
  journal = {Physical Review E},
  volume  = {61},
  number  = {1},
  pages   = {R45--R48},
  year    = {2000},
  doi     = {10.1103/PhysRevE.61.R45},
}

@article{Ramos1999,
  author  = {Ramos, Antonio and Morgan, Hywel and Green, Nicolas G. and Castellanos, Antonio},
  title   = {{AC} electric-field-induced fluid flow in microelectrodes},
  journal = {Journal of Colloid and Interface Science},
  volume  = {217},
  number  = {2},
  pages   = {420--422},
  year    = {1999},
  doi     = {10.1006/jcis.1999.6346},
}

@article{Bazant2004,
  title={Diffuse-charge dynamics in electrochemical systems},
  author={Bazant, M.Z. and Thornton, K. and Ajdari, A.},
  journal={{Physical Review E}},
  volume={70},
  number={2},
  pages={021506},
  year={2004},
  publisher={APS}
}

@article{Rubinstein2009,
  title={Reexamination of electrodiffusion time scales},
  author={Rubinstein, I. and Zaltzman, B. and Futerman, A. and Gitis, V. and Nikonenko, V.},
  journal={{Physical Review E}},
  volume={79},
  number={2},
  pages={021506},
  year={2009},
  publisher={APS}
}

@article{Sanabria2006,
  title={Relaxation processes due to the electrode-electrolyte interface in ionic solutions},
  author={Sanabria, H. and Miller{,} Jr., J.H.},
  journal={Physical Review E—Statistical, Nonlinear, and Soft Matter Physics},
  volume={74},
  number={5},
  pages={051505},
  year={2006},
  publisher={APS}
}

@article{Solis2023,
  title={Electrical properties of tissues from a microscopic model of confined electrolytes},
  author={Solis, F.J. and Jadhao, V.},
  journal={Physics in Medicine \& Biology},
  volume={68},
  number={10},
  pages={105017},
  year={2023},
  publisher={IOP Publishing}
}
\end{document}


\preprint{APS/123-QED}
\title{Universal behavior in diffuse charge dynamics with patterned electrodes - Supplementary Material
}

\author{E. Krucker-Velasquez}
\affiliation{Center for Computation and Theory of Soft Materials, Robert R. McCormick School of Engineering and Applied Science, Northwestern University, Evanston IL 60208 USA
}

\author{E. Kirkinis}
\affiliation{Center for Computation and Theory of Soft Materials, Robert R. McCormick School of Engineering and Applied Science,  Northwestern University, Evanston IL 60208 USA
}
\affiliation{Department of Materials Science \& Engineering, Northwestern University, Evanston IL 60208 USA}

\author{M. Olvera de la Cruz}
\affiliation{Center for Computation and Theory of Soft Materials, Robert R. McCormick School of Engineering and Applied Science,  Northwestern University, Evanston IL 60208 USA
}
\affiliation{Department of Physics and Astronomy, Northwestern University, Evanston, IL 60208 USA}
\affiliation{Department of Materials Science \& Engineering, Northwestern University, Evanston IL 60208 USA}
\date{}
\begin{abstract}
In this Addendum we provide closed analytic forms for bulk charge $\rho$ and electric potential $\phi$ for the cell displayed in Fig. 1(a) of the main Letter. We derive Eq. (2) of the Letter for the impedance with non-uniformly charged walls and establish universality by showing its independence with respect to Neumann or Dirichlet boundary conditions. We provide an analysis 
for the onset of frequency dispersion at thin Debye layers leading to Eq. (4) of the Letter. 
We show that voltage- and charge-driven cells share the same quality factor, leading to universality for their common natural frequency.
Finally, we 
examine the range of validity of our theoretical formulation.
\end{abstract}

\maketitle

\section{\label{sec: charge}Charge and potential for confined electrolytes}
We consider a $1:1$ electrolyte with ionic concentrations $c_\pm=c_\infty+\delta c_\pm$, $\delta c_\pm$ denoting small perturbations about the uniform bulk concentration $c_\infty$.
The charge density is $\rho=e(\delta c_+-\delta c_-)$, while the salt density, expressed in charge units, is $s=e(c_++c_-)\simeq2ec_\infty$ to leading order. Here, $e$ is the elementary charge.
{
The validity of this approximation is examined in Appendix \ref{sec: validity}, following the line of thought developed in
 \cite{Shrestha2025}. \\

\indent The electric potential $\phi$ is related to the charge density by Poisson's equation,
\be \label{phit}
\nabla^2 \phi = -\frac{\rho}{\epsilon},
\ee
where $\epsilon$ is the liquid permittivity. For equal ionic diffusivities, the charge density evolves according to
\be \label{rhotgeneral}
\frac{\partial \rho}{\partial t}
=
D\left[
\nabla^2 \rho
+
\frac{e}{k_BT}\nabla\cdot(s\nabla\phi)
\right],
\ee
where $D$ is the common ionic diffusion coefficient 
\footnote{Unequal ionic diffusivities can introduce additional features in Nyquist plots and enhance transport velocities in binary electrolytes \cite{Henrique2025,*Shrestha2025b}.}
and $k_BT$ is the thermal energy. Substituting the leading-order salt density, $s\simeq 2ec_\infty$, and using Eq.~\eqref{phit} gives the Debye-Falkenhagen equation
\citep{Bazant2004},
\be \label{rhot}
\frac{\partial\rho }{\partial t}
=
D\left[\nabla^2\rho-\kappa^2\rho\right],
\ee
where
\be \label{kappa}
\kappa
=
\left(\frac{2e^2c_\infty}{\epsilon k_BT}\right)^{1/2}
\ee
is the inverse Debye length.
\subsection{\label{sec: bcs}Boundary conditions}
{ 
We consider Neumann boundary conditions for the Poisson equation \rr{phit}, 
at a solid wall of surface charge $\sigma$ 
where
\be \label{N1}
 \hat{\mathbf{n}} \cdot \nabla  \phi  = -\frac{\sigma}{\epsilon}, \quad \textrm{at a charged wall}
\ee
with the unit normal vector $\hat{\mathbf{n}} $ pointing into the liquid. Blocking electrodes (current of bulk charge $\rho$ vanishes at wall) are described by 
\be \label{jn}
 \mathbf{j}\cdot \hat{\mathbf{n}} \equiv  - D\left[ \nabla \rho + \frac{e}{k_BT} s \nabla \phi  \right] \cdot \hat{\mathbf{n}}=0
 , \quad \textrm{at a charged wall}.
\ee
This also leads to a boundary condition for the bulk charge $\rho$
\be \label{N2}
 \hat{\mathbf{n}} \cdot \nabla  \rho = \kappa^2 \sigma, \quad \textrm{at a charged wall}
\ee
replacing \rr{jn}. 
 
\subsection{\label{sec: channel}Potential and bulk charge in a charging cell}
The surface charges are applied on the cell walls at $z=\pm h$ leading to  
boundary conditions for the electric potential and charge
\be \label{phirhochannel}
\partial_z \phi  = -   \frac{\sigma(x, t)}{\epsilon}, \quad \partial_z \rho  = \kappa^2 \sigma(x, t), 
\ee
with $\sigma(x,t) = \sigma_0 e^{ -i \omega t}\cos kx$, as described in Fig. 1(a) of the Letter.

Let
\be \label{kdelta2}
P \equiv P_1+iP_2 = \sqrt{\frac{i\omega}{D}  -\kappa^2-k^2}.
\ee

where
\be \label{P12}
P_{1,2} = \frac{1}{\sqrt{2}} \sqrt{\sqrt{\left(\kappa^{2}+k^2\right)^{2}+\frac{\omega^{2}}{\mathit{D}^{2}}}\mp (\kappa^{2}+k^2)},
\ee
and the minus/plus sign corresponds to the real/imaginary part of $P$. 
Assuming $\rho = \rho(z) e^{ -i \omega t}\cos kx$ and subject  to the 
boundary condition \rr{phirhochannel} the bulk charge satisfies $ \rho_{zz} + \left[i\omega/D -\kappa^2-k^2 \right] \rho =0$.
To avoid clutter, we introduce the notation
$\rho_z\equiv \partial_z \rho$ etc. Solving this equation, 
the bulk charge distribution becomes
\be \label{rho}
\rho(x,z,t) = \frac{\sigma_0\kappa^2\sin Pz}{P\cos  Ph} e^{ -i\omega t}\cos kx,  
\ee
where $P$ is the  complex wavenumber defined in \rr{kdelta2}. 

Similarly, assuming $\phi = \phi(z) e^{ -i\omega t}\cos kx$ and subject  to the 
boundary condition \rr{phirhochannel} the electric potential becomes
\begin{align}
\phi(x,z,t) =&- \frac{\sigma_0}{\epsilon} \left( 1 + \frac{\kappa^2}{P^2 } \right) \frac{\sinh kz}{k \cosh kh} e^{ -i\omega t}\cos kx
\nonumber 
\\
&+\frac{1}{\epsilon (P^2 + k^2) } \rho, 
\label{phi1}
\end{align}
with $\rho$ given by \rr{rho}.

The transient response is determined by averaging \rr{phit} and \rr{rhot} over the width $2h$ of the cell
and employing boundary conditions \rr{phirhochannel}. This gives
\be
\langle \rho \rangle = \rho_\infty e^{-D(\kappa^2+k^2)t}, \quad 
\langle \phi \rangle = -
\frac{\langle \rho \rangle}{\epsilon k^2},
\ee
which defines the relaxation time 
\be
\tau_D = \frac{1}{D(\kappa^2+k^2)}\sim \frac{1}{D\kappa^2}
\ee
employed in the main body of the Letter.

\section{Frequency-dependent complex Impedance}
The above formulation can be thought of arising due to a transverse oscillating electric field. It can become somewhat simplified by establishing 
correspondence with \citep{Solis2023}. 
Let 
\be
E_B = \frac{\sigma_0}{\epsilon} \frac{i\omega}{DP^2}.
\ee
Then
\be \label{phi2}
\phi (z) = \left[ - \frac{\sinh kz}{k \cosh kh}  + \frac{D\kappa^2}{i\omega P} \frac{\sin Pz}{\cos  Ph} \right] E_B. 
\ee
Correspondence with the notation of \citep{Solis2023} can be established with 
\be
\omega \rightarrow -\omega
\quad \textrm{and} \quad
P=iq
\ee
where $P_2>0$ so $q_1>0$. 

\subsection{\label{sec: current}Current expressions}
Following \citet{Solis2023} we employ the notations
\be
\mathbf{j}, \quad \mathbf{j}_d, \quad \textrm{and} \quad \mathbf{J}
\ee
to denote the ionic current, displacement current and electromagnetic current, respectively,
related by 
\be \label{EMcurrent}
\mathbf{J} = \mathbf{j} + \mathbf{j}_d
\ee
where 
\be
 j = - D\left[ \frac{\partial \rho}{\partial z} + \epsilon \kappa^2 \frac{\partial \phi}{\partial z} \right] \quad \textrm{and} \quad
 j_d = - \epsilon \partial_t  \frac{\partial \phi}{\partial z}
\ee
and $\phi$ is given by \rr{phi2}.

The electromagnetic current \rr{EMcurrent} (ionic + displacement) in the electrolyte is given by 
\be \label{J1}
{J}(\omega, z) = - D\left[ \frac{\partial \rho}{\partial z} + \epsilon \kappa^2 \frac{\partial \phi}{\partial z} \right] - \epsilon \partial_t  \frac{\partial \phi}{\partial z}.
\ee
Since $\phi = \phi(z) e^{ -i \omega t}\cos kx$, $\phi (z) = - \frac{\sinh kz}{k \cosh kh} E_B + \frac{1}{\epsilon (P^2 + k^2)} \rho $ (from \rr{rho}-\rr{phi2}) and 
$D\kappa^2 -i\omega = -D(P^2+k^2)$, current \rr{J1} becomes 
\begin{align} \label{elcond}
J(\omega, z)& = - D\left[ \frac{\partial \rho}{\partial z} - \epsilon (P^2 + k^2) \frac{\partial \phi}{\partial z} \right] \nonumber \\
&\equiv - \epsilon D (P^2+k^2) E_B \frac{\cosh kz}{\cosh kh}.
\end{align}
{Thus, the electromagnetic current in the electrolyte region is \emph{not} spatially uniform} (as is the case when $k\equiv0$).

The potential difference in the electrolyte region is 
\be \label{phiS}
\Delta \phi_S = \phi(h) - \phi(-h) = -2 h \left[  \frac{\sinh kh}{kh \cosh kh} - \frac{D\kappa^2}{i\omega } \frac{{\tan Ph}}{Ph}  \right] E_B
\ee
where we employed the electric potential expression \rr{phi2}. 

\begin{figure*}
\vspace{-5pt}
\begin{center}
\includegraphics[height=2.6in,width=5.6in,angle=0]{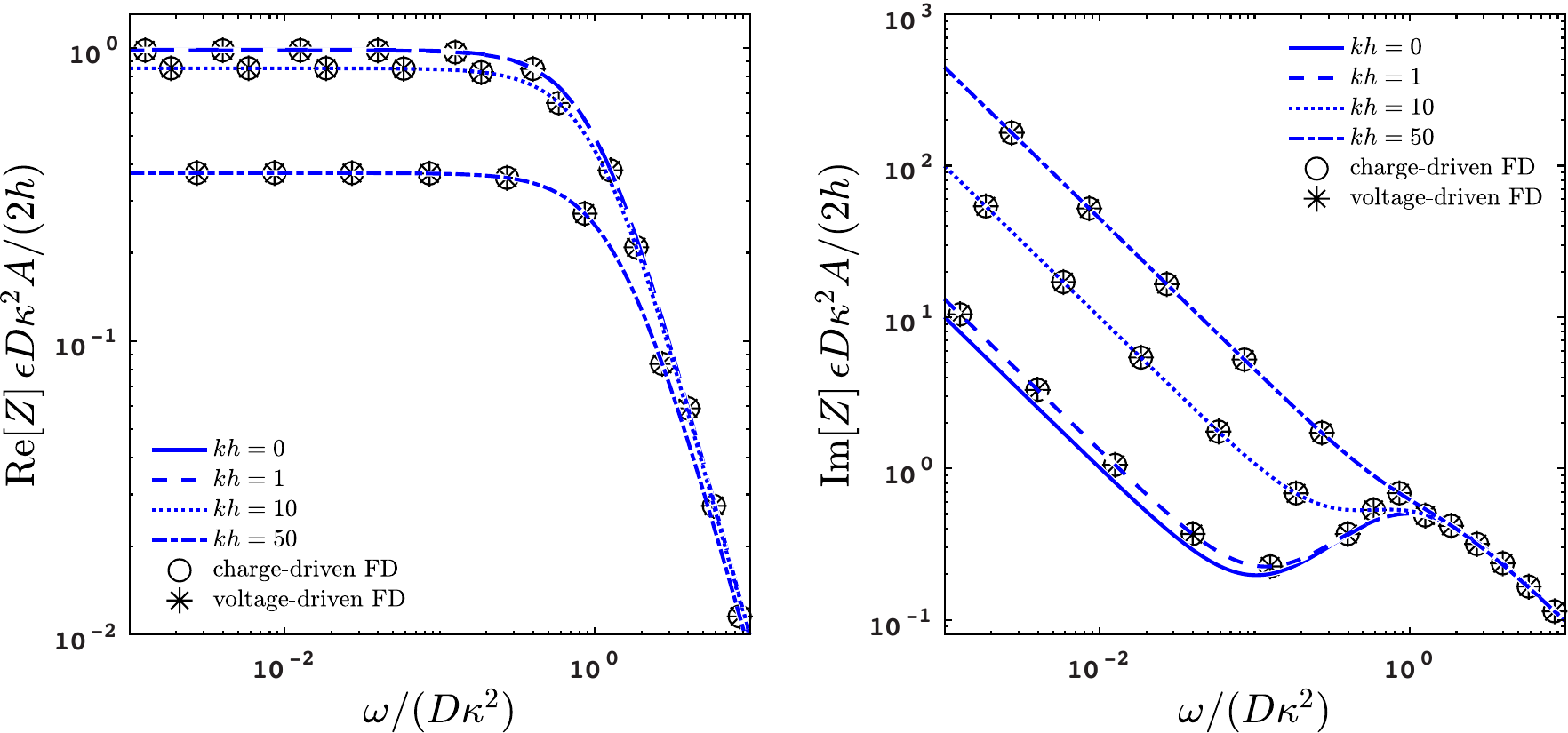}
\vspace{-15pt}
\end{center}
\caption{Real part (left panel) and imaginary part (right panel) of the impedance Eq. (2) of the Letter, for various values of the wall charge modulation wavenumber $k$ and cell semi-width $h$ as a function of dimensionless frequency $\omega/(D\kappa^2)$. Open circles and stars correspond to values obtained from finite difference simulations of the charge- and voltage-driven cells. 
\label{reZimZomega}  }
\vspace{0pt}
\end{figure*}

\subsection{Impedance}
The impedance expression is given by \cite{Solis2023}
\be \label{Zdef}
Z = \frac{V(\omega)}{A I(\omega)},
\ee
where $V(\omega)$ is the amplitude of the potential drop in one period of the system, $I(\omega)$ a suitable measure of the current  see Fig. 1(a) of the Letter for the definition).

As in \citep{Solis2023} we define 
\be \label{Vomega}
V(\omega) =- \Delta \phi_S \equiv 2h \left[ \frac{\sinh kh}{kh \cosh kh} - \frac{D\kappa^2}{i\omega } \frac{{\tan Ph}}{Ph}   \right] E_B
\ee
where we employed \rr{phiS} for the potential. 

The electromagnetic current \rr{elcond} in the electrolyte is spatially-dependent. We can define the current $I(\omega)$ to be
the average of the electromagnetic current in the electrolyte region, thus 
\be \label{Iomega}
I(\omega) \equiv \langle J \rangle =  -\epsilon D(P^2 + k^2) E_B \frac{\tanh kh}{kh},
\ee
employing  \rr{elcond}. 
Expression \rr{Zdef} leads to the impedance relation Eq. (2) of the Letter by invoking expressions \rr{Vomega} and \rr{Iomega} above. 

The factor $\frac{\tanh(kh)}{kh}$ describes the redistribution of current produced by the lateral wall pattern and $\frac{\tan(Ph)}{Ph}$ contains the frequency-dependent diffuse-charge response. Since $P$ is complex and depends on $\omega$, the latter term produces both the dissipative and reactive components of the impedance. The absence of $\sigma_0$ from Eq. (2) of the Letter is a consequence of linear response: both $V$ and $I$ are proportional to the forcing amplitude, and their ratio therefore is independent of it. 

For completeness and to accompany the Nyquist plot in Fig. 1(c) of the Letter, we display the real and imaginary parts of the wall-charge modulated impedance Eq. (2) of the Letter vs. frequency, for various values of the electrode charge modulation wavenumber $k$ in Fig. \ref{reZimZomega}.

The impedance Eq. (2) of the Letter displays a self-similar behavior. For the dimensions of the parameters $\kappa, k, D, \omega,h$ define a matrix of rank two. Thus, there are three dimensionless 
combinations that can formed \citep{Panton1996,*Bluman1989}
\be \label{Zss}
Z =\frac{2h}{\epsilon D\kappa^2A} \mathscr{Z}
\left(\frac{\omega}{D\kappa^2}, \kappa h,kh \right),
\ee
$\mathscr{Z}$ is a dimensionless nonlinear function of the three independent dimensionless parameters. Thus, Fig. 1(c) of the Letter displays the real and imaginary parts of $\mathscr{Z}$. 
\newcommand{\taud}{\tau_\mathrm{D}}
\newcommand{\taurck}{\tau_{\mathrm{RC},k}}
\newcommand{\kappad}{\kappa}
\newcommand{\lambdad}{\lambda_\mathrm{D}}
\newcommand{\bigO}{\mathcal{O}}
\newcommand{\eps}{\varepsilon}
\newcommand{\tZ}{\tilde{Z}}
\section{\label{sec:sqrt6} Derivation of the dispersion-onset frequency $\omega_{\star} \approx \sqrt{6}\omega_0$}

Starting from the \emph{dimensionless} impedance, $\tZ= Z\epsilon D \kappad^2 A/ 2h$,
\begin{equation}
    \tZ(\Omega) = \frac{ 1 - \frac{kh}{\tanh (kh)} \frac{\tan (Ph)}{\ii \Omega P h } }{ 1-\ii \Omega}\, ;
    \label{eq:Zfirst}
\end{equation}
where $\Omega =\omega \taud = \omega/( D\kappa^2 )$ is the frequency scaled on the inverse Debye relaxation time. The
 substitution $P^2 + k^2 = \kappad^2(\ii \Omega -1)$ was used in the denominator of eq. ~\eqref{eq:Zfirst}. Because $\tan(\ii x)=\ii \tanh x$, we can make the substitution $\tan(Ph)/Ph=\tanh(y)/y$ where $y=\kappad h\sqrt{1-\ii \Omega +k^2 \lambdad^2}$, such that

\begin{equation}
    \tZ(\Omega) = \frac{ 1 - \frac{kh}{\tanh (kh)}\frac{\tanh (y)}{\ii \Omega y } }{ 1-\ii \Omega}\, ;
\end{equation}
 in the thin double layer limit, $y\sim \kappad h \gg1$ and $\tanh(y)\approx 1$, leading to
\begin{equation}
     \tZ\approx\frac{1}{(1-\ii\Omega)} + \frac{\ii \frac{kh}{\tanh (kh)}}{ \Omega \kappad h(1+k^2 \lambdad^2)^{1/2}( 1 -\ii \Omega/(1+k^2 \lambdad^2) )^{1/2}(1-\ii\Omega)}\;.
\end{equation}
The first term in the impedance describes the bulk response, whereas the second describes the diffuse-layer contribution. Assuming that $k^2\lambdad^2\ll1$, this simplifies to
\begin{equation}
     Z\approx\frac{1}{(1-\ii\Omega)} + \frac{\ii \varepsilon^2}{ \Omega (1-\ii \Omega)^{3/2}}\;.
\end{equation}
where $\varepsilon^2 = \taud/\taurck \ll1$ such that the Debye and pattern-dependent relaxation times are well separated. At low frequency, their imaginary parts give $ \operatorname{Im}\tZ \approx   \Omega+\varepsilon^2/\Omega $
The bulk contribution increases with frequency, while the diffuse-layer contribution decreases. Their crossover therefore occurs when  $\Omega=O(\varepsilon)$. To set the frequency scale in the crossover region and retain the bulk and diffuse layer contributions at leading order, we measure the frequency relative to the geometric-mean frequency, $1/\sqrt{\taud \taurck}$, via $u$:
\begin{equation}
    \Omega=\varepsilon u,
    \qquad
    u=\frac{\Omega}{\varepsilon}
     =\omega\sqrt{\taud\taurck}.
    \label{eq:crossover_frequency}
\end{equation}
With this rescaling, the thin-layer impedance becomes
\begin{equation}
    \tZ(\varepsilon u)
    \simeq
    \frac{1}{1-\ii\varepsilon u}
    +
    \frac{\ii\varepsilon}
    {u(1-\ii\varepsilon u)^{3/2}}.
    \label{eq:impedance_crossover}
\end{equation}
Expanding the factors $(1-\ii\varepsilon u)^{-1}$ and
$(1-\ii\varepsilon u)^{-3/2}$ in powers of $\varepsilon$ at fixed
$u$, and writing $\tZ = \tZ^\prime +\ii \tZ^{\prime \prime}$
\begin{align}
    \tZ^\prime(u)
    &=
    1-\varepsilon^2\left(\frac32+u^2\right)
    +\mathcal{O}(\varepsilon^4),
    \label{eq:impedance_real_expansion}
    \\
    \tZ^{\prime\prime}(u)
    &=
    \varepsilon\left(u+\frac1u\right)
    -\varepsilon^3\left(u^3+\frac{15}{8}u\right)
    +\mathcal{O}(\varepsilon^5).
    \label{eq:impedance_imag_expansion}
\end{align}
\begin{figure}[ht!]
\centering
\includegraphics[width=0.3\textwidth]{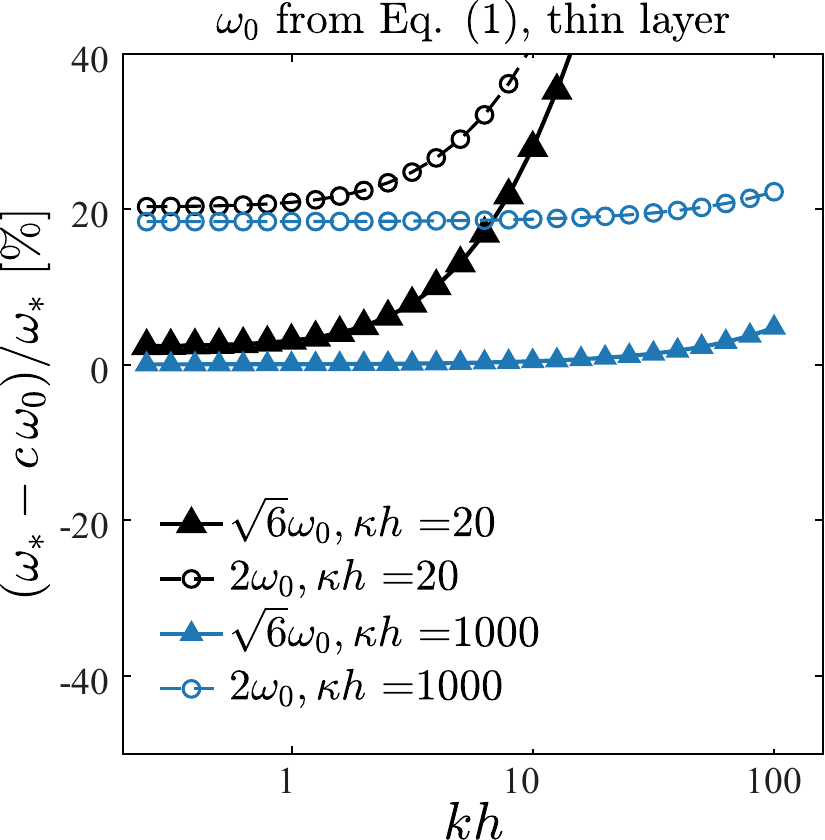} 
\caption{Relative error (as percentage) for the estimated dispersion onset as a function of $\kappa h$. Open circled markers correspond to the empirical approximation $\omega_\star = 2\omega_0$, and filled triangled markers correspond to the derived $\omega_\star = \sqrt{6}\omega_0$, valid in the thin double-layer limit.}
\label{fig:errors}  
\end{figure}
We identify the onset of dispersion with the inflection point which occurs when the signed curvature changes signs
\begin{equation}
    Z^\prime_u Z^{\prime\prime}_{uu} - Z^{\prime\prime}_{u}  Z^\prime_{uu} = -\frac{4\eps^3}{u^2} +2\eps^3(1-\frac{1}{u^2})=0 \;.
    \label{eq:curvature_zero}
\end{equation}
The condition in eq. ~\eqref{eq:curvature_zero} is satisfied when
\begin{equation}
    \Omega^\star = \sqrt{3}\eps = \sqrt{\frac{3 \taud}{\taurck}}
\end{equation}
with $\omega = \Omega/\taud$
\begin{equation}
    \omega^\star = \frac{\sqrt{3}}{\sqrt{\taud \taurck}}=\sqrt{3\times 2 }\omega_0\approx 2.45\omega_0
\end{equation}

\section{\label{sec: QF_V}Quality factor for the voltage driven cell}
\begin{align}
       \hrho &= A\sinh pz \;, \nonumber \\
       \hphi &= -\frac{A}{\epsilon(P^{2}+k^{2})}  \sinh p z - \frac{\ii \omega}{D\kappa^2}p \cosh(ph)\frac{\sinh kz}{k\cosh kh}
       \label{eq:hrho_hphi}
\end{align}
For voltage driven:
\begin{equation}
A = \frac{\epsilon V_0(\kappa^2 - \ii \omega/D)}{\sinh ph - \frac{\ii \omega}{D\kappa^2 }\frac{p}{k}\tanh kh\cosh ph }
\label{eq:Avolt}
\end{equation}
let 
\begin{equation}
\hat{g} = \hrho + \epsilon \kappa^2 \hphi 
\end{equation}
such that the ionic current is $\mbf{i}=-D\grad\hat{g}$. 

\begin{equation}
\mathcal{B}_k[f,g]\equiv\frac{1}{2h}\int_{-h}^{h}\left(f_z g_z^{*}+k^2 f g^{*}\right)\,dz .
\label{eq:Einner}
\end{equation}
\begin{align}\overline{\mathcal{E}}&=\frac{\epsilon}{4}
\mathcal{B}_k[\widehat{\phi},\widehat{\phi}], \nonumber \\
\overline{\mathcal{P}}&=\frac{D}{2}\left[\Re\mathcal{B}_k[\widehat{\rho},\widehat{\phi}]+\epsilon\kappa^2\mathcal{B}_k[\widehat{\phi},\widehat{\phi}]
    \right]\nonumber \\
    &=\frac{D}{2}\Re \Bk[\hg,\hphi]=\frac{D}{2\epsilon\kappa^2}\Bk[\hg,\hg],
\label{eq:energy_power_compact_Q}
\end{align}
such that 
\begin{equation}
    Q=\omega \frac{\Eb}{\Pb}\;.
    \label{eq:quality_compact}
\end{equation}
Integrating by parts with the no-peneration condition $\hg^{\prime}(\pm h) = 0$ and $\hg^{\prime\prime} - k^2\hg=-(\ii \omega D)\hrho$ gives $2h \Bk[\hg,\hrho] = (\ii\omega/D)\int|\hrho|^2 \mrm{d}z$, purely imaginary, so $\Re \Bk[\hg,\hg]/(\epsilon\kappa^2)$. The same integration by parts applied to $\Bk[\hg,\hphi]$ gives:
\begin{align}
        \Pb &= \frac{\omega \epsilon}{2 h }\Im[\hphi^\prime(h)\hphi(h)]= \frac{\omega \epsilon V_0}{2 h }\Im \hphi^\prime(h) \nonumber \\
        &=-\frac{\omega V_0}{2 h \kappa^2}\Im \hrho^\prime(h)
\end{align}
Substituting eq. ~\eqref{eq:hrho_hphi} with ~\eqref{eq:Avolt} in eq. ~\eqref{eq:energy_power_compact_Q} we find that
\begin{equation}
\Eb_V = \frac{\epsilon V_0^2}{4}\frac{\Bk[T,T]}{|T(h)|^2},\quad \Pb_V = \frac{\epsilon \omega^2 V_0^2}{2D\kappa^2}\frac{\Bk[S,S]}{|T(h)|^2}
\end{equation}
where 
\begin{align}
        S(z) &= \sinh pz - p\cosh ph \frac{\sinh kz}{ k \cosh kh}\; , \nonumber\\
         T(z) &= \sinh pz - \frac{\ii \omega}{D \kappa^2 }p\cosh ph \frac{\sinh kz}{k \cosh kh}\;,
\end{align}
for the charge drive the same expression holds with the prefactor replaced as $V_0^2/|T(h)|^2\to\kappa^4\sigma_0^2(|1/(\kappa^2-\ii\omega/D)|^2)/|p\cosh ph|^2$. The amplitude cancels in the ratio:

\begin{equation}
    Q_V(\omega) = Q_\sigma (\omega) =\frac{D\kappa^2}{2\omega}\frac{\Bk[T,T]}{\Bk[S,S]}\;.
\end{equation}
Consequently, they share the same damping $2r=D\kappa^2$ and the same natural frequency. 
\newcommand{\hphip}{\hphi_{\mrm{pen}}}
\newcommand{\hphis}{\hphi_{\mrm{scr}}}
\section{\label{sec:screenvpenetrating} Energy of the screened and penetrating field}
Here we compare the energy of the screened against the penetrating field. The expression for the potential in eq. ~\eqref{phi2} can be rewritten in terms of the contributions from the penetrating, $\hphip(z)$, and screened, $\hphis(z)$, potentials:
\begin{equation}
    \hphi(z) = -E_B\left( \frac{\sinh(kz)}{k\cosh(kh)} + \frac{\ii}{\Omega} \frac{\sinh(pz)}{p\cosh(ph)}\right) = \hphip + \hphis\;
\end{equation}
where the penentrating field is given by the Laplace component $ \sinh(kz)/(k\cosh(kh))$, and the screened component is controlled by $ \sinh(pz)/(p\cosh(ph))$. Following the definition of the electric field energy in eq. ~\eqref{eq:energy_power_compact_Q}, the ratio of the energies of the two field components to leading order is
\begin{equation}
    \frac{\Eb_{\mrm{pen}}}{\Eb_{\mrm{src}}}\approx 2(\omega \taud)^2 \frac{\tanh(kh)}{k\lambdad} = \left( \frac{\omega}{\omega_0}\right)^2\;.
\end{equation}
\section{\label{sec: validity}Validity of the Debye-Falkenhagen approximation}
We examine the conditions under which the reduction of the charge evolution equation \rr{rhot} and 
associated boundary condition is valid. The requirement is that $\rho \ll 2 ec_\infty$. 
We will show that this requirement translates to 
\be \label{cond0}
Re\left\{  \frac{\sigma_0 \kappa^2 \sin Pz}{P \cos Ph}    \right\} \leq
 \frac{\epsilon E\kappa^2 \coth (P_2h)}{\left[ (\kappa^2+k^2)^2 + \left( \frac{\omega}{D}\right)^2 \right]^{\frac{1}{4}}}  \ll 2 ec_\infty,
\ee
where we replaced the wall charge by the nominal electric field $E = \sigma_0/\epsilon$. 
Letting $P= R e^{i\Theta}$ where $R = \left[ (\kappa^2+k^2)^2 + \left( \frac{\omega}{D}\right)^2 \right]^{1/4}$ and $P = P_1 + i P_2$ (defined in \rr{P12}), it is sufficient to calculate the modulus of $\frac{\sin Pz}{ \cos Ph}$.
Thus, 
\begin{widetext}
\begin{align}
\left|\frac{ \sin Pz}{ \cos Ph} \right|^2 &= \frac{\sin^2 P_1z \cosh^2 P_2z + \cos^2P_1z \sinh^2P_2z}{\cos^2 P_1h \cosh^2 P_2h + \sin^2P_1h \sinh^2P_2h} = 
 \frac{\sin^2 P_1z (1 + \sinh^2 P_2z) + (1 - \sin^2P_1z) \sinh^2P_2z}{(1- \sin^2 P_1h) \cosh^2 P_2h + \sin^2P_1h ( \cosh^2P_2h-1)} 
 \nonumber \\
 &= 
 \frac{\sin^2 P_1z + \sinh^2P_2z}{\cosh^2P_2h-\sin^2 P_1h} \leq 
\frac{\sinh^2 P_2z + 1}{\cosh^2P_2h-1} = \frac{\cosh^2 P_2z}{\sinh^2 P_2h} \leq \coth^2 P_2h.
\end{align}
\end{widetext}

Eq. \rr{cond0} can be  reexpressed more clearly by eliminating $ec_\infty$  in favor of $\kappa$ through \rr{kappa}
\be \label{cond0a}
f(\kappa,  E) \equiv \frac{eE \coth P_2h }{k_BT  \left[ (\kappa^2+k^2)^2 + \left( \frac{\omega}{D}\right)^2 \right]^{1/4} } \ll 1.
\ee
This condition is identical to the one derived in \cite{Shrestha2025} for a boundary-guided electrokinetic configuration with traveling wave charges based on the work of \citet{Ehrlich1982}. 
The bound \rr{cond0a} differs from the one derived in the standard 
Debye-H\"uckel approximation (see Eq. \rr{condition1}) in that the inverse Debye length $\kappa$ in the latter is replaced here by the renormalized inverse Debye wavelength $R$. As can be seen in the left-most panel of figure \ref{impedance_validity} the conclusions of our formulation significantly improve compared to their Debye-H\"uckel counterparts (right-most panel of figure \ref{impedance_validity})
even for moderate frequencies $\omega\sim 10^2$ rad/sec. 

In the left two panels of figure \ref{impedance_validity} we display three-dimensional contours of the function $f(\kappa, E) $ in \rr{cond0a} for $f= 1, 0.1, 0.01$ and $0.001$, $k=10^5\; \textrm{m}^{-1}$, and $D=10^{-9}\; \textrm{m}^{2}/\textrm{sec}$ at room temperature showing that
the range of validity of the approximation increases as the frequency $\omega$ increases.

To make a connection with the Debye-H\"{u}ckel approximation, consider the small Debye length limit Eq.  \rr{cond0} leading to
$\sigma_0 \kappa \ll 2 c_\infty$. Eliminating $c_\infty$ (or taking the large $\kappa$ limit in \rr{cond0a}) we obtain 
\be \label{condition1}
\frac{eE}{\kappa k_BT } \ll 1
\ee
which is analogous to the Debye-H\"{u}ckel approximation where $e\phi \ll k_BT$ \citep[p. 10.22]{Melcher1981}, if the 
characteristic length-scale of the system is the Debye length. 
The right-most panel of 
figure \ref{impedance_validity} displayes three-dimensional contours of the function $\frac{eE}{\kappa k_BT }$ at room temperature. In conclusion, the 
range of validity of the Debye-Falkenhagen approximation leading to the bound \rr{cond0a} and displayed in the two leftmost panels of figure \ref{impedance_validity}, is superior to its Debye-H\"{u}ckel counterpart (displayed in the right-most panel of Fig. \ref{impedance_validity}). 

\begin{figure*}
\vspace{-5pt}
\begin{center}
\includegraphics[height=2in,width=7.4in,angle=0]{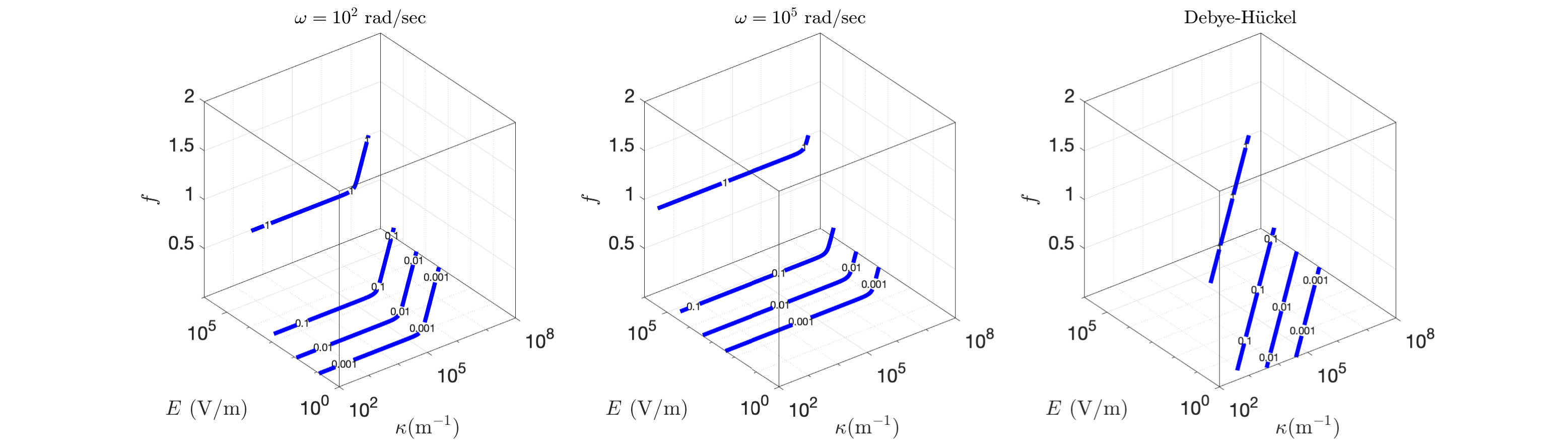}
\vspace{-10pt}
\end{center}
\caption{Validity of the Debye-Falkenhagen equation determined by requiring the function $f(\kappa, E) $ in \rr{cond0a} to be $\ll 1$. In the left two panels we display contours of $f(\kappa, E) $ in \rr{cond0a} compared with the Debye-H\"uckel
approximation (contours of the function $\frac{eE}{\kappa k_BT }$ in  \rr{condition1} in the right-most panel). 
The plateau in the left two panels is due to the frequency renormalization $R$ of the Debye wavenumber 
in the denominator of Eq. \rr{cond0a}. 
Thus, 
the range of validity of the approximation increases as the frequency $\omega$ (or the ratio $\omega/D$) increases.
Here $k=10^5\; \textrm{m}^{-1}$, and $D=10^{-9}\; \textrm{m}^{2}/\textrm{sec}$ at room temperature. 
\label{impedance_validity}  }
\vspace{0pt}
\end{figure*}

\nocite{Henrique2025,Shrestha2025b}
\bibliography{biblio}